\documentclass[aps,pra,floatfix,twocolumn,tightenlines,amsmath,amssymb]{revtex4}

\usepackage{graphicx}
\usepackage{amsmath}
\usepackage{amssymb}
\usepackage{epstopdf}
\usepackage{color}

\newcommand{\ket}[1] {|#1 \rangle}

\newcommand*{\rom}[1]{\expandafter\@slowromancap\romannumeral #1@}

\begin{document}

\title{Tunnelling statistics for analysis of spin readout fidelity}
\author{S. K. Gorman, Y. He, M. G. House, J. G. Keizer, D. Keith, L. Fricke, S. J. Hile, M. A. Broome, M. Y. Simmons}
\affiliation{Centre of Excellence for Quantum Computation and Communication Technology, School of Physics, University of New South Wales, Sydney, New South Wales 2052, Australia}
\date{\today}

\begin{abstract}
We investigate spin and charge dynamics of a quantum dot of phosphorus atoms coupled to a radio-frequency single-electron transistor (rf-SET) using full counting statistics. We show how the magnetic field plays a role in determining the bunching or anti-bunching tunnelling statistics of the donor dot and SET system. Using the counting statistics we show how to determine the lowest magnetic field where spin-readout is possible. We then show how such a measurement can be used to investigate and optimise single electron spin-readout fidelity.
\end{abstract}

\maketitle

\section{Introduction}
\label{sec:intro}

Single shot electron spin-readout is crucial for scalable quantum computation in silicon~\cite{Hille1500707,ogorman2016}. The single electron transistor (SET) has proven to be a highly sensitive electron charge detector in recent years~\cite{Knobel2003,1347-4065-52-4S-04CJ01,RevModPhys.85.961} and is routinely used to perform high fidelity electron spin-readout when operated in DC mode~\cite{PhysRevB.81.161308,A.2016,PhysRevLett.115.166806,morello2010}. The SET can also be operated in AC mode using rf-reflectometry, which has been shown to increase detection bandwidths and give larger signal-to-noise ratios (SNR)~\cite{Schoelkopf1238,:/content/aip/journal/apl/74/26/10.1063/1.123258,:/content/aip/journal/apl/86/14/10.1063/1.1897423,:/content/aip/journal/apl/107/9/10.1063/1.4929827}. However, it is not known how a rf-driving field will affect the fidelity of electron spin-readout or if electron spin-readout is even possible in devices where the electron is tunnel coupled to a rf-SET. To investigate the combined rf-SET and electron system for single shot spin-readout, we examine the statistical properties of electrons tunnelling between a single donor dot (DD) comprised of $\sim$5 P atoms and a rf-SET.

Electron spin-readout is governed by spin-selective tunnelling processes of an electron from a DD to an electron reservoir~\cite{elzerman2004}. In particular, the tunnel out rates of the electron spin-up and spin-down states from the DD to the reservoir need to be vastly different to ensure high-fidelity spin-to-charge conversion~\cite{PhysRevLett.115.166806}. If the tunnel rates are too similar they cannot be attributed to the correct qubit state. Importantly, in such a system the tunnelling statistics of electrons to and from a reservoir can provide a vast amount of information about the underlying physical processes for the coupled DD-SET  system~\cite{Lu2003,hanson2007,PhysRevB.91.235413,PhysRevB.94.054314}. This information can in turn be used to optimise the spin-readout fidelity using full counting statistics (FCS)~\cite{PhysRevB.67.085316,Nazarov2003,1367-2630-16-11-113061}. In addition, FCS can be used to investigate shot noise~\cite{gustavsson2006,0953-8984-20-45-454204}, non-Markovian effects~\cite{PhysRevLett.100.150601,Flindt23062009,flindt2010} and electron-electron interactions~\cite{PhysRevLett.96.026805,doi:10.1063/1.3430000} that are difficult to obtain from transport measurements alone.

Full counting statistics involves counting the number of tunnel events, $n$ of an electron typically between a reservoir and electronic state such as a quantum dot within a time window, $\tau$~\cite{Wang20159}. By repeatedly counting the tunnel events over many multiples of $\tau$ a number distribution of tunnel events, $p(n)$ can be obtained~\cite{gustavsson2006}. The resulting distribution can be completely described by a set of cumulants, $\kappa_i$ derived from the natural logarithm of the moment-generating function of $p(n)$. The cumulants represent different statistical properties about the number distribution, in which, $\kappa_1$, $\kappa_2$, and $\kappa_3$ are the mean, variance, and skewness, respectively~\cite{bruderer2014,PhysRevB.91.235413}. Knowledge of the tunnelling statistics can then be used to optimise the time and energy detuning for electron spin-readout since they rely on the tunnelling of electrons from the DD to the reservoir.

In this work, we show by analysing the random telegraph signal (RTS) produced from the DD electron tunnel events, how the system varies under different magnetic field and rf-power conditions. The paper is laid out in the following sections. In Sec.~\ref{sec:exp}  we describe the operation of the device and outline the measurement of RTS traces. We then derive the first few cumulants in terms of the electronic tunnel rates in the system in Sec.~\ref{sec:rts}. We investigate the dependence of FCS on magnetic field and rf-power in Sec.~\ref{sec:Bfield}, and in more detail, the low, high and intermediate magnetic field regimes in Sec.~\ref{sec:Blow}-\ref{sec:Binter}. In Sec.~\ref{sec:readout} we present a short overview of fidelity analysis of spin-readout and discuss how to optimise the readout time in Sec.~\ref{sec:Optimtime}, as well as the rf-power of the rf-SET in Sec.~\ref{sec:Optimpower}. Finally, in Sec.~\ref{sec:disc} we summarise the results and describe potential future extensions to this work.

\section{Device characterisation}
\label{sec:exp}

The device was patterned using scanning tunnelling microscopy (STM) - hydrogen lithography to selectively remove a hydrogen mask and subsequently dosed with phosphine to incorporate phosphorus donors~\cite{simmons2008}, see Fig.~\ref{fig:intro}a. The device was mounted on a printed circuit board with a rf-tank circuit that had a resonant frequency of 228.6 MHz ($L{=}1200$~nH and parasitic capacitance, $C_p{\sim}0.4$~pF), a matching capacitor of $C_m{=}39$~pF and Q-factor $\sim$150 attached to the source contact, while the drain contact was grounded~\cite{PhysRevApplied.6.044016}. The amplitude of the reflected signal was monitored throughout the experiment and a variable attenuator was used to adjust the input rf-power driving the SET. Whilst two ${\sim}5$ P DDs were patterned in the device, we concentrate on the right DD in this paper since we are only interested in the single electron dynamics between the DD and rf-SET. Figure~\ref{fig:intro}b shows a charge transition between the DD and rf-SET with the detuning, $\epsilon$ between the DD and SET along the white arrow.

To acquire the RTS traces we position the chemical potential of the DD such that an electron can tunnel to the rf-SET. We then monitor the rf-SET for $100$~s before shifting the chemical potential along the detuning direction shown by the white arrow in Fig.~\ref{fig:intro}b (see Appendix A for details of the FCS analysis). The reflected rf-amplitude RTS traces are digitised with a 500 kHz sample rate with an example trace shown in Fig.~\ref{fig:intro}c. The low level (blue) of the RTS trace corresponds to the DD having an extra electron, whereas the high level (yellow) indicates when an electron has tunnelled off the DD to the SET reservoir. We set a threshold level, shown as a red dashed line in Fig.~\ref{fig:intro}c that distinguishes between the DD charge states, `0' (yellow) and `1' (blue). The number of electrons on the SET and DD is given by ($N_{\textnormal{SET}}$,$N_{\textnormal{DD}}$) in the figure and do not represent absolute numbers since we have not depleted the DD for this experiment.

\subsection{Random telegraph signal analysis}
\label{sec:rts} 

\begin{figure}
\begin{center}
\includegraphics[width=1\columnwidth]{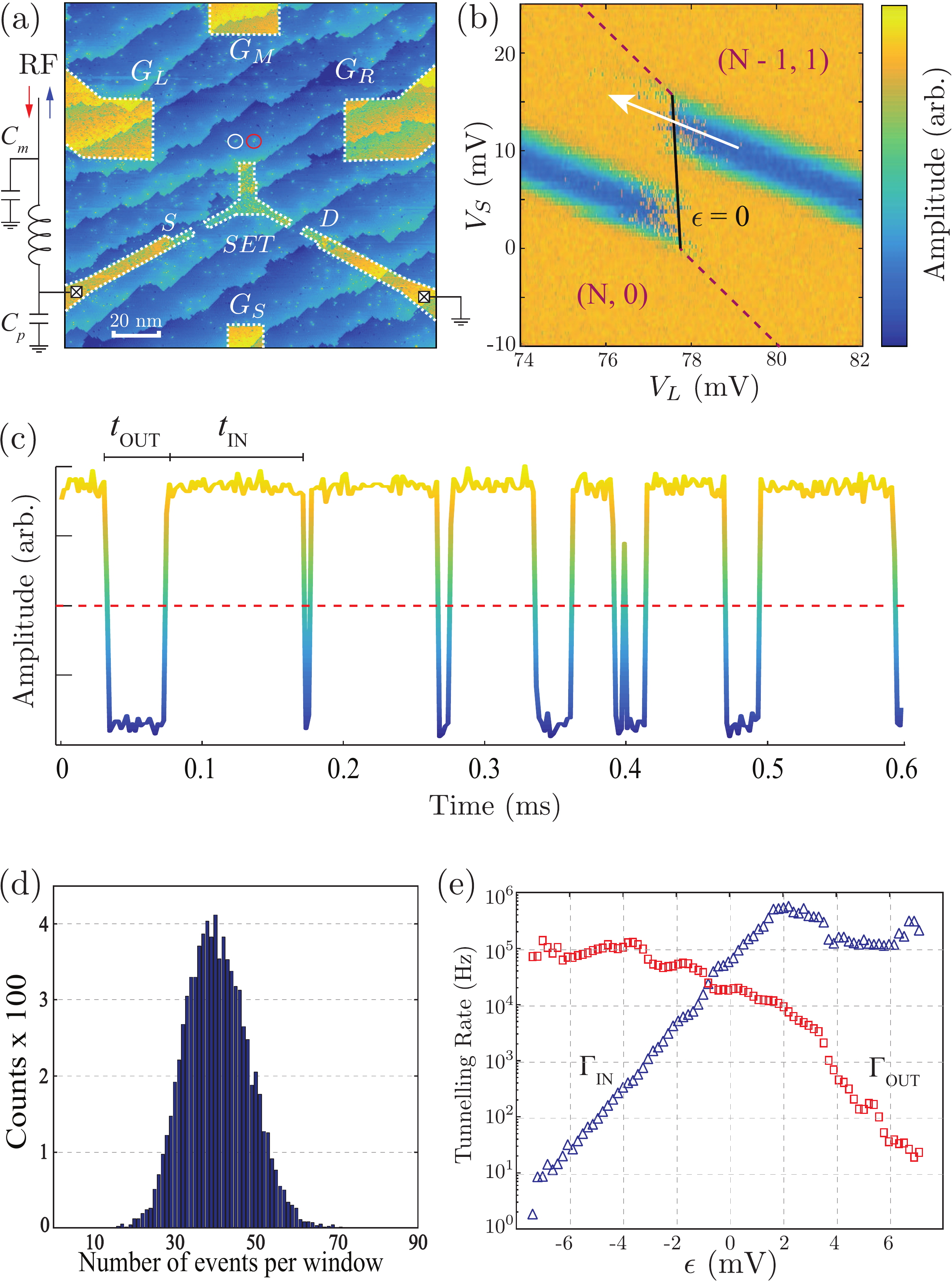}
\end{center}
\caption{{\bf Full counting statistics of a few donor quantum dot coupled to an in-plane rf-SET.} (a) A STM-micrograph of the device investigated. Two DDs were patterned in the device; however, in the paper we only study the right DD (red circle). There are three control gates for the DDs $\{G_L,G_M,G_R\}$ and one for the rf-SET, $G_S$. The rf-tank circuit is attached to the source contact and the drain is grounded. The tank circuit is characterised by $L{=}1200$~nH, a parasitic capacitance, $C_p{\sim}0.4$~pF, and a matching capacitor, $C_m{=}39$~pF. A variable attenuator is used to tune the rf-power reaching the device. (b) An anti-crossing between the DD and rf-SET at $B{=}0$~T in the reflected amplitude of the rf-signal showing relative electron numbers on SET and DD ($n_{\textnormal{SET}}$,$n_{\textnormal{DD}}$). The detuning axis, $\epsilon$ is shown by the white arrow. (c) An illustrative RTS trace taken by measuring the reflected amplitude of the rf-SET near $\epsilon{=}0$ in (b). (d) The resulting distribution, $p(n)$ after using FCS to analysis the RTS trace in (c) at $B{=}0$~T. The histogram shows a mean of a ${\sim}40$ with a variance of ${\sim}20$. The distribution is positively skewed, that is $SF{>}0$. (e) The measured tunnel rates as a function of detuning at $B{=}0$~T showing the Fermi distribution about the Fermi level of the rf-SET.}
\label{fig:intro}
\end{figure}

To encapsulate the complete dynamics of the system, we consider the system evolving under the Liouville equation, assuming the Born-Markov approximation,
\begin{equation}
\frac{d \rho}{d t} = \mathcal{L} \rho,
\end{equation}
where $\rho$ is the density operator and $\mathcal{L}$ is the generator of the system, which includes both coherent and incoherent tunnelling processes. The cumulants for a given generator $\mathcal{L}$ can be found by using FCS to analyse the RTS. Here, we use the recently proposed characteristic polynomial approach~\cite{bruderer2014} which links the generator to the cumulants of the number distribution of tunnel events, $p(n)$, see Fig~\ref{fig:intro}d where the tunnel out events are used to generate the distribution. In addition, we extract the distribution of waiting times of the `0' and `1' states from which we can determine the tunnel rates as a function of detuning as shown in Fig.~\ref{fig:intro}e at $B{=}0$~T.

\begin{figure}
\begin{center}
\includegraphics[width=1\columnwidth]{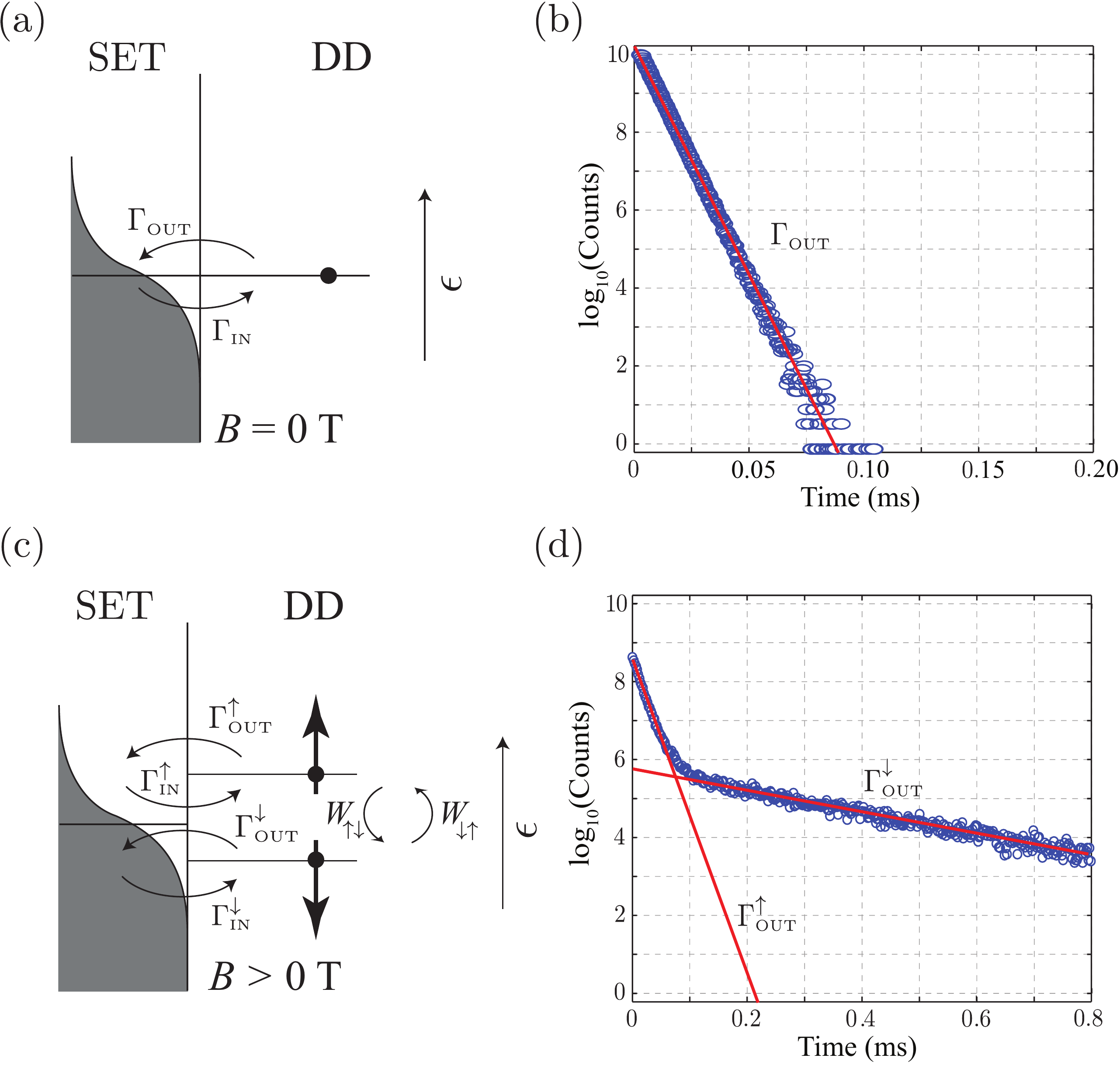}
\end{center}
\caption{{\bf The effect of an applied magnetic field to the distribution of waiting times.}  A schematic of the detuning and spin states, $\ket{{\uparrow}}$ and $\ket{{\downarrow}}$ at (a) $B{=}0$~T and (b) $B{>}0$~T showing the individual tunnel rates between the DD and SET with thermal broadening. (c) At $B{=}0$~T the spin states are degenerate and only a single exponential decay in the distribution of waiting times is observed. (d) At $B{>}0$~T the spin states are split by the Zeeman energy causing two distinct tunnel out rates of the DD to the rf-SET. As a result, the distribution of waiting times shows a double exponential decay for $\Gamma^{\uparrow}_{\textnormal{OUT}}$ and $\Gamma^{\downarrow}_{\textnormal{OUT}}$.}
\label{fig:waittimes}
\end{figure}

For degenerate spin states ($B{=}0$~T, see Fig.~\ref{fig:waittimes}a), the system has only two distinguishable states, ${\ket{0}}$ when there is no electron on the DD  and $\ket{1}$ when there is one electron on the DD. The generator, $\mathcal{L}$ in the basis $\{{\ket{0}},{\ket{1}}\}$, contains the tunnel rates of the electron between the DD and reservoir,
\begin{equation}
\mathcal{L}_{0} = \begin{pmatrix}
-\Gamma_{\textnormal{IN}} & \Gamma_{\textnormal{OUT}}\\
\Gamma_{\textnormal{IN}} & -\Gamma_{\textnormal{OUT}}\\
\end{pmatrix},
\label{eqn:Lodef}
\end{equation}
where $\Gamma_{\textnormal{IN}}$ ($\Gamma_{\textnormal{OUT}}$) is the tunnel rate from the SET to DD (DD to SET) shown in the distribution of waiting times in Fig.~\ref{fig:waittimes}b. To perform FCS, we introduce a counting field, $\xi$ over the transition that is measured by examining the RTS traces, see Fig.~\ref{fig:intro}c. This transforms $\mathcal{L}_0 \rightarrow \mathcal{L}^{\xi}_0$, which is given by
\begin{equation}
\mathcal{L}^{\xi}_0 = \begin{pmatrix}
-\Gamma_{\textnormal{IN}} & e^{\xi}\Gamma_{\textnormal{OUT}}\\
\Gamma_{\textnormal{IN}} & -\Gamma_{\textnormal{OUT}}\\
\end{pmatrix},
\end{equation}
where the counting is performed over the tunnel out events of the DD to the rf-SET ($\ket{1}{\rightarrow}\ket{0}$). The choice of tunnel in or out events does not affect the FCS analysis and the same $p(n)$ can be obtained by counting over the tunnel in events from the rf-SET to the DD.

To calculate the cumulants of $\mathcal{L}^{\xi}_0$ we use the recently proposed characteristic polynomial approach to counting statistics~\cite{bruderer2014}. This method uses the characteristic polynomial, $P^{\xi}(z){=}\det{[z\mathcal{I} - \mathcal{L}^{\xi}]}$ of the generator (where $z$ is a placeholder variable and $\mathcal{I}$ is the identity matrix) to find the cumulants rather than finding the smallest eigenvalue of the generator~\cite{PhysRevB.67.085316}. The notable benefit of the characteristic polynomial approach is that analytical expressions for the cumulants can always be obtained since it is not necessary to find the eigenvalues of the generator (the roots of $P^{\xi}(z)$)~\cite{bruderer2014}. In addition, statistical tests of the system dimension can be derived and the measured cumulants can be inverted to determine an unknown generator~\cite{bruderer2014}. Therefore, the characteristic polynomial allows for more information to be gained from the counting statistics compared to the standard approach~\cite{PhysRevB.67.085316}.

In general, the characteristic polynomial, $P^{\xi}(z)$ is related to the cumulants of the generator through the total derivative of $P^{\xi}[\lambda(\xi)]$ with respect to the counting field, $\xi$~\cite{bruderer2014},
\begin{equation}
\frac{d^l P^{\xi}[\lambda(\xi)]}{d \xi^l}\Bigg|_{\xi = 0} = 0 \hspace{10pt} l \geq 1,
\label{eq:dp}
\end{equation} 
where $\lambda(\xi)$ is the smallest eigenvalue of the generator. Evaluating Eq.~\ref{eq:dp} for $l{=}\{1,2,3\}$ and taking into account the relations, $\kappa_i{=}\partial_{\xi}^i\lambda(\xi)|_{\xi{=}0}$ and $\lambda(\xi)|_{\xi{=}0}{=}0$, we can solve for the cumulants, $\kappa_1$, $\kappa_2$, and $\kappa_3$~\cite{bruderer2014},
\begin{equation}
\kappa_1 = -\frac{a_0'}{a_1},
\label{eq:k1}
\end{equation}
\begin{equation}
\kappa_2 = -\frac{1}{a_1}(a_0' + 2a_1'\kappa_1 + 2a_2\kappa^2_1),
\label{eq:k2}
\end{equation}
\begin{multline}
\kappa_3 = -\frac{1}{a_1}(a_0' + 3a_1'\kappa_1 + 6a_2\kappa^2_1 + 6a_3\kappa^3_1 \\+ 3a_1'\kappa_2 + 6a_2\kappa_1\kappa_2),
\label{eq:k3}
\end{multline}
where $a_n$ is the $n^{\textnormal{th}}$ coefficient of $z$ in the characteristic polynomial. Similarly, $a_n'$ is the derivative of the $n^{th}$ coefficient of $z$ with respect to $\xi$ in the limit that $\xi{\rightarrow}0$. Using Eq.~\ref{eq:k1}, \ref{eq:k2}, and \ref{eq:k3} we can readily find the analytical expressions for the first three cumulants from the coefficients of the characteristic polynomial.

The characteristic polynomial of the $\mathcal{L}^{\xi}_0$ in the case of degenerate spin states has the form,
\begin{equation}
P^{\xi}_0(z) = z^2 + (\Gamma_{\textnormal{IN}} + \Gamma_{\textnormal{OUT}})z + \Gamma_{\textnormal{OUT}}\Gamma_{\textnormal{IN}}(1 - e^{\xi}),
\end{equation}
Substituting in the coefficients of $P^{\xi}_0(z)$ gives,
\begin{equation}
\kappa_1 = \frac{\Gamma_{\textnormal{IN}}\Gamma_{\textnormal{OUT}}}{\Gamma_{\textnormal{IN}} + \Gamma_{\textnormal{OUT}}},
\end{equation}
\begin{equation}
\kappa_2 = \kappa_1\frac{\Gamma^2_{\textnormal{IN}} + \Gamma^2_{\textnormal{OUT}}}{(\Gamma_{\textnormal{IN}} + \Gamma_{\textnormal{OUT}})^2},
\end{equation}
\begin{multline}
\kappa_3 = \kappa_1 \times \\
\frac{\Gamma_{\textnormal{IN}}^4-2\Gamma_{\textnormal{IN}}^3\Gamma_{\textnormal{OUT}}+6\Gamma_{\textnormal{IN}}^2\Gamma_{\textnormal{OUT}}^2-2\Gamma_{\textnormal{IN}}\Gamma_{\textnormal{OUT}}^3+\Gamma_{\textnormal{OUT}}^4}{(\Gamma_{\textnormal{IN}} + \Gamma_{\textnormal{OUT}})^4}.
\end{multline}

For energy selective electron spin-readout, the electron Zeeman split energy levels, $\ket{{\uparrow}}$ and $\ket{{\downarrow}}$ must have an energy separation, $g \mu_B B{>}k_B T$, where $g \mu_B B$ is the Zeeman energy for magnetic field strength $B$ and $k_B T$ is the thermal energy at temperature, $T$. The spin split levels are then positioned with the Fermi level of a reservoir between them, such that only $\ket{{\uparrow}}$ can tunnel out and $\ket{{\downarrow}}$ can tunnel in. However, due to temperature broadening, there is a finite probability that the electrons can tunnel back and forth between the reservoir and DD indefinitely.

When a magnetic field is applied, the spin states become non-degenerate and we must now consider a three state system, see Fig.~\ref{fig:waittimes}c. In this case, each spin state has distinct dynamics due to their different chemical potential with respect to the Fermi level of the SET, resulting in different tunnel rates, $\{\Gamma^{\downarrow}_{\textnormal{IN}}, \Gamma^{\downarrow}_{\textnormal{OUT}}, \Gamma^{\uparrow}_{\textnormal{IN}}, \Gamma^{\uparrow}_{\textnormal{OUT}} \}$ as well as inter-spin relaxation rates, $W_{\uparrow \downarrow}$ and $W_{\downarrow \uparrow}$. This added complexity significantly changes the cumulants of the system and hence six tunnel rates are now required to describe the DD-SET tunnelling. This is clearly demonstrated by examining the distribution of waiting times in Fig.~\ref{fig:waittimes}d where two exponential decays are observed corresponding to the individual spin tunnel rates. In the limit that the inter-spin relaxation rates are much smaller than the DD-SET tunnel rates, $\kappa_1$ is given by,
\begin{equation}
\kappa_1 = \frac{\Gamma^{\downarrow}_{\textnormal{IN}}\Gamma^{\uparrow}_{\textnormal{OUT}}(\Gamma^{\uparrow}_{\textnormal{IN}} + \Gamma^{\downarrow}_{\textnormal{IN}})}{\Gamma^{\downarrow}_{\textnormal{IN}}\Gamma^{\uparrow}_{\textnormal{OUT}} + \Gamma^{\downarrow}_{\textnormal{OUT}}(\Gamma^{\uparrow}_{\textnormal{IN}} + \Gamma^{\uparrow}_{\textnormal{OUT}})}.
\end{equation}
This is the case for most donor systems with long spin relaxation times, $T_1{>}1$~s at $B{=}2.5$~T. The higher order cumulants can be calculated in an equivalent manner to the $B{=}0$ T case; however, their general analytical form is too large to quote~\cite{PhysRevB.71.161301}.

The normalised second cumulant, known as the the Fano factor (\textit{FF}) defined as $\textrm{\textit{FF}}{=}\kappa_2/\kappa_1$ is a useful quantity when investigating the system dynamics since it gives information about the temporal distribution of the tunnel events. That is, for tunnel events that are evenly separated in time (anti-bunching), $\textrm{\textit{FF}}{<}1$ and for tunnel events that are clustered with long periods of no tunnelling (bunching), $\textrm{\textit{FF}}{>}1$. We also make use of the normalised skewness, $SF{=}\kappa_3/\kappa_1$. While $FF$ must be positive, $SF$ can range from $-\infty$ to $\infty$. For $SF{<}0$, $p(n)$ extends further in $n$ values less than $\kappa_1$, that is the distribution is negatively skewed. Conversely, for $SF{>}0$, $p(n)$ has more values larger than $\kappa_1$ meaning the distribution is now positively skewed. In particular, a Gaussian distribution is described by $SF{=}0$ or more precisely, $\kappa_{i>2}{=}0$. We make use of $SF$ in Sec.~\ref{sec:Binter} to determine the lowest magnetic field where the spin states are distinguishable.

\begin{figure*}[t!]
\includegraphics[width=1\textwidth]{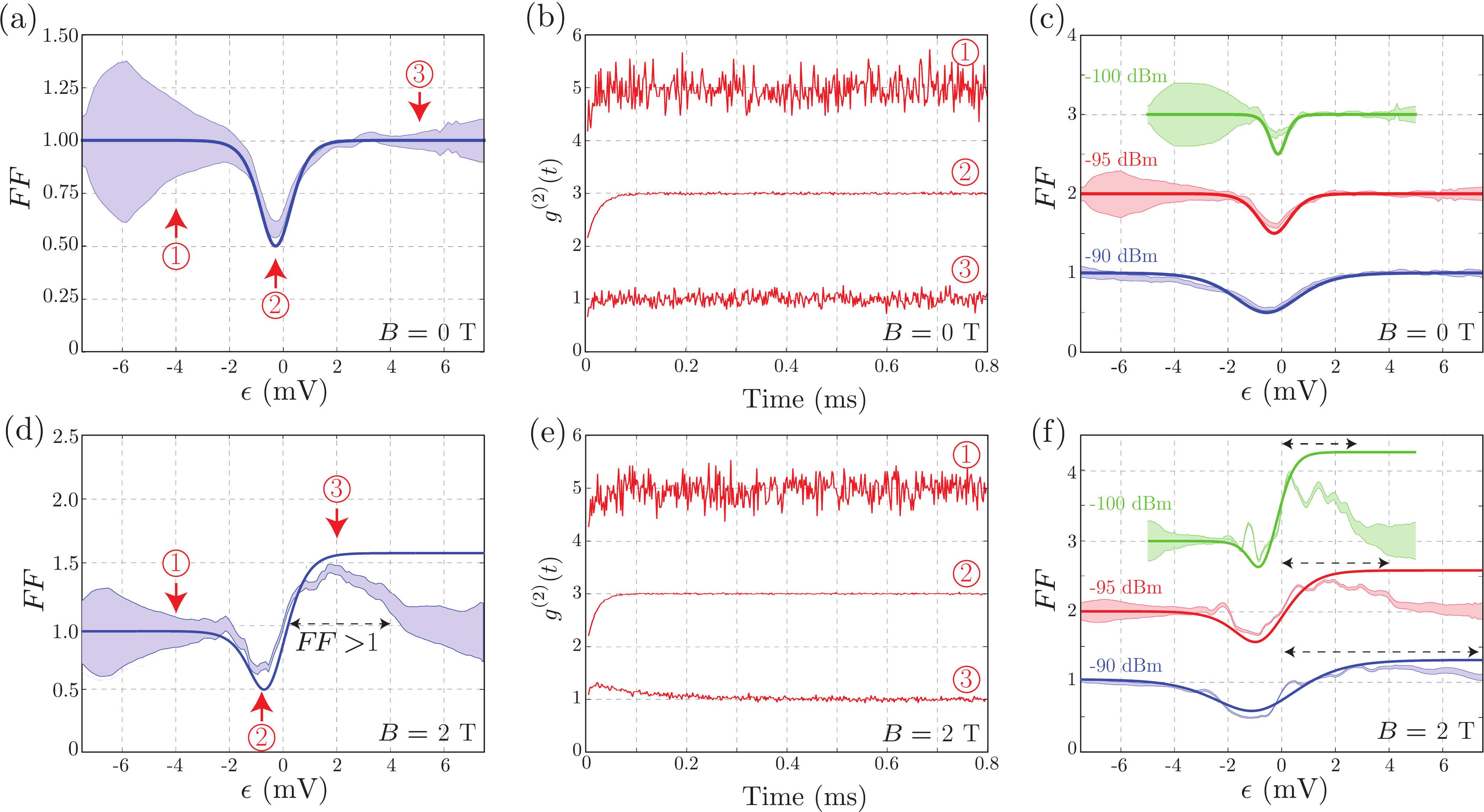}
\caption{{\bf Second-order correlation function and power-dependence of the Fano factor as a function of detuning.} (a) Fano factor, $\textrm{\textit{FF}}{=}\hat{\kappa}_2/\hat{\kappa}_1$ as a function of detuning, $\epsilon$ between the DD and rf-SET at $B{=}0$ T and -95~dBm. There is a dip to $\textrm{\textit{FF}}{\sim}0.5$ near zero detuning, indicating anti-bunching of electrons tunnelling between the DD and the rf-SET. The shaded region is the confidence interval of the experimental data and the solid line is a fit assuming a Fermi distribution of the SET. (b) Second-order correlation function, $g^{(2)}(t)$ at the three different detuning positions (offset by 2) for $B{=}0$ T labelled in (a). All detuning positions show anti-bunching, $g^{(2)}(t){\leq}1$ confirming the \textit{FF} measurement in (a). The dips near $t{=}0$ for \raisebox{.5pt}{\textcircled{\raisebox{-.9pt} {1}}} and \raisebox{.5pt}{\textcircled{\raisebox{-.9pt} {3}}} are due to the limited bandwidth of the rf-SET. (c) The power dependence of the \textit{FF} for three different rf-powers (-90, -95, and -100 dBm) applied to the SET (offset by 1). As the power is increased the \textit{FF} broadens as the result of a larger effective electron temperature. (d) \textit{FF} at $B{=}2$ T. There is now a peak in the \textit{FF} above 1, which indicates bunching of electron tunnel events due to the different tunnel rates between the Zeeman split spin states. (e) $g^{(2)}(t)$ at three different detuning positions for $B{=}2$ T (offset by 2). For position \raisebox{.5pt}{\textcircled{\raisebox{-.9pt} {2}}} anti-bunching is also observed as the $\ket{{\downarrow}}$ state is aligned with the rf-SET Fermi level and therefore electrons can tunnel back and forth to the SET. At position \raisebox{.5pt}{\textcircled{\raisebox{-.9pt} {3}}} there is clear evidence of bunching of electrons, $g^{(2)}(t){>}1$, confirming the \textit{FF} measurements in (d). Again, the sharp dip near $t{=}0$ is due to the limited bandwidth of the rf-SET. (f) The power dependence of \textit{FF} at $B{=}2$ T (offset by 1). The peak in the \textit{FF} becomes less pronounced as rf-power increases. This is a result of the increasing electron temperature since the height of the peak depends on the difference in tunnel out rate between $\ket{{\downarrow}}$ and $\ket{{\uparrow}}$ states. The width of the peak (dashed line) increases since there is a larger detuning range over which sufficient tunnelling statistics can be obtained.}
\label{fig:ff_g2_c3}
\end{figure*}

\section{Magnetic field dependence}
\label{sec:Bfield}

To fully understand the tunnelling dynamics of the DD-SET  system, we consider three distinct magnetic field regimes---low ($g \mu_B B{<}k_B T$) and high ($g \mu_B B{>}k_B T$) magnetic field---where the system can be described by a two-state ($\{{\ket{0}}, {\ket{1}}\}$) and three-state ($\{{\ket{0}}, {\ket{\downarrow}},{\ket{\uparrow}}\}$) system, respectively and the intermediate magnetic field ($g \mu_B B{\sim}k_B T$) case. We then use a statistical test to determine the magnetic field for which the system can no longer be described by a two-state model.

\subsection{Low Magnetic field}
\label{sec:Blow}

First, we consider the system at $B{=}0$ T for which the spin states are degenerate. The tunnelling of the electrons produces a RTS trace such as shown in Fig.~\ref{fig:intro}c corresponding to the two charge states of the DD, $\ket{0}$ (high level) and $\ket{1}$ (low level).

The shaded bands in Fig.~\ref{fig:ff_g2_c3}a shows the \textit{FF} of the tunnel events of the DD to the SET ($\hat{\kappa}_i$ are cumulants determined from the experimental data, see Appendix~\ref{app:calc}) as a function of detuning, $\epsilon$ between the DD and rf-SET (see arrow in Fig.1 b). There is a single dip near $\epsilon{=}0$ (position \raisebox{.5pt}{\textcircled{\raisebox{-.9pt} {2}}}) that has a minimum of $\sim$0.55. This is an indication of electron anti-bunching~\cite{gustavsson2006}, in which the electron tunnel out events are evenly spaced out in time. This is due to the Fermionic nature of the electron such that only one can occupy a specific DD energy level at a time. It is worth noting that the \textit{FF} does not reach 0.5 since the tunnel rates are extremely sensitive at $\epsilon{=}0$ and small electrical noise fluctuations can change them significantly. As a result, on average the tunnel rates are not exactly equal at $\epsilon{=}0$ and there is some additional variance in the counting statistics introduced from the noise in the system. The \textit{FF} then approaches 1 for $\epsilon{\ll}0$ (position \raisebox{.5pt}{\textcircled{\raisebox{-.9pt} {1}}}) and $\epsilon{\gg}0$ (position \raisebox{.5pt}{\textcircled{\raisebox{-.9pt} {3}}}) where the electron becomes Coulomb blockaded and cannot tunnel between the DD and SET. The \textit{FF} agrees very well with theoretical calculations (solid lines) where only an effective temperature is assumed, as is standard practice for a DC SET. For example, at -100~dBm, the effective power-broadened temperature is ${\sim}1.4$~K (see Appendix~\ref{app:temp} for details on the temperature calculation). Since the tunnelling statistics can be described by a simple effective temperature broadening, in the same manner as a DC SET, we conclude that the rf-SET is suitable for single shot electron spin-readout. Note that below we perform the same experiment with much lower rf-driving powers, and hence a lower power broadened temperature.

Although the \textit{FF} can distinguish between the overall behaviour of the tunnelling dynamics, to examine the temporal correlations of tunnel events we make use of the second-order correlation function, $g^{(2)}(t)$~\cite{PhysRevB.85.165417},
\begin{equation}
g^{(2)}(t) = \frac{\langle\langle\mathcal{J} e^{\mathcal{L}t} \mathcal{J}\rangle\rangle}{\langle\langle\mathcal{J}\rangle\rangle^2},
\label{eq:g2calc}
\end{equation}
where $\mathcal{J}{=}\frac{d}{d\xi}\mathcal{L}(\xi)|_{\xi{=}0}$ is the jump operator for the counting field, $\xi$ and $\langle\langle \dots \rangle\rangle$ indicates the steady state average. The $g^{(2)}(t)$ can be used to distinguish between anti-bunching ($g^{(2)}(t){<}1$) and bunching ($g^{(2)}(t){>}1$) tunnel events. Experimentally, $g^{(2)}(t)$ is calculated by building a histogram of the times, $t$ between every pair of tunnel out events in the RTS trace. We note by definition, $g^{(2)}(0){=}0$ since electrons are Fermions, that is, we cannot detect individual tunnel events that are not separated in time~\cite{PhysRevB.85.165417}.

The second-order correlation function is shown in Fig.~\ref{fig:ff_g2_c3}b at the 3 different detuning positions, marked \{\raisebox{.5pt}{\textcircled{\raisebox{-.9pt} {1}}},\raisebox{.5pt}{\textcircled{\raisebox{-.9pt} {2}}},\raisebox{.5pt}{\textcircled{\raisebox{-.9pt} {3}}}\} in Fig.2a. At $\epsilon{\approx}0$ (point \raisebox{.5pt}{\textcircled{\raisebox{-.9pt} {2}}}) $g^{(2)}(t){<}1$ for $t{<}0.1$~ms indicating that electron anti-bunching is observed at these short timescales. For large detuning (\raisebox{.5pt}{\textcircled{\raisebox{-.9pt} {1}}} and \raisebox{.5pt}{\textcircled{\raisebox{-.9pt} {3}}}) where the DD is in Coulomb blockade, $g^{(2)}(t){=}1$ since the tunnel events are not correlated in time. This confirms the observation of the $FF$ dip in Fig.~\ref{fig:ff_g2_c3}a.

Finally, in this low field regime we investigate the counting statistics as a function of the applied rf-power to examine the effect of any artificial broadening due to rf-driving of the SET. Any broadening due to excessive rf-power is relevant when considering electron spin-readout fidelities, which are strongly reduced at high electron temperatures. Figure~\ref{fig:ff_g2_c3}c shows the \textit{FF} as a function of detuning for three rf-powers. We observe that increased rf-power broadens the \textit{FF} dip, indicating that the higher power causes a higher effective temperature of the SET. However, it does not significantly effect the counting statistics since the tunnelling dynamics can still be explained by the simple DC reservoir model (solid lines). Therefore, although the rf-driving of the SET does not change the tunnelling dynamics, the rf-power needs to be chosen carefully as not to power broaden the SET which will ultimately decrease the electron spin-readout fidelity, in particular as $k_B T{\rightarrow}g \mu_B B$. 

\subsection{High magnetic field}
\label{sec:Bhigh}

We now examine the high magnetic field case where the electron spin state can be read out since the spin split levels are sufficiently distinct to allow spin-to-charge conversion~\cite{buch2013}. Therefore, it is important to characterise the dynamics of the non-degenerate spin states using the rf-SET to determine any detrimental effects that may affect single shot spin-readout.

At large magnetic fields $g \mu_B B{>}k_B T$ the dynamics can no longer be explained by a two level system. The Zeeman split levels now have their own dynamics and the generator must describe a three-state system. The generator, $\mathcal{L}_B$ in the basis $\{{\ket{0}},{\ket{{\downarrow}}},{\ket{{\uparrow}}}\}$ of the DD electron is given by,
\begin{equation}
\mathcal{L}_B = \begin{pmatrix}
-\Gamma^{\uparrow}_{\textnormal{IN}} -\Gamma^{\downarrow}_{\textnormal{IN}} & \Gamma^{\downarrow}_{\textnormal{OUT}} & \Gamma^{\uparrow}_{\textnormal{OUT}}\\
\Gamma^{\downarrow}_{\textnormal{IN}} & -\Gamma^{\downarrow}_{\textnormal{OUT}} - W_{\downarrow\uparrow} & W_{\uparrow\downarrow}\\
\Gamma^{\uparrow}_{\textnormal{IN}} & W_{\downarrow\uparrow} & -\Gamma^{\uparrow}_{\textnormal{OUT}} - W_{\uparrow\downarrow}\\
\end{pmatrix},
\label{eq:threestate}
\end{equation}
where $\Gamma^i_n$ are the tunnel rates for the individual spin states in and out of the DD and $W_{\uparrow \downarrow}$ ($W_{\downarrow \uparrow}$) is the relaxation rate from $\ket{{\uparrow}}{\rightarrow}\ket{{\downarrow}}$ ($\ket{{\downarrow}}{\rightarrow}\ket{{\uparrow}}$).

The measured \textit{FF} as a function of detuning at $B{=}2$~T is shown in Fig.~\ref{fig:ff_g2_c3}d. There is a dip near $\epsilon{=}0$ (position \raisebox{.5pt}{\textcircled{\raisebox{-.9pt} {2}}}) as seen in the low magnetic field case, indicative of anti-bunching. The \textit{FF} then rises above 1 ($\epsilon{>}0$) for a length of detuning before returning to 1. This arises due to the bunching of the electron tunnel events due to the different tunnel rates of the Zeeman split states. The width in detuning for which $\textrm{\textit{FF}}{>}1$, shown by the dashed lines, is related to the temperature of the system and Zeeman splitting of the spin states. The discrepancy between the theoretical curve and data at far positive detuning is due to the finite window size, $\tau$ in our analysis. If $\tau{\sim}1/\Gamma^i_n$ then FCS breaks down and the number distribution cannot be described accurately. As a result, the distribution becomes Poissonian, $\kappa_{i>1}{=}\kappa_1$ and hence \textit{FF}${\rightarrow}1$.

The detuning for which the \textit{FF} rises above 1 shows where bunching of the $\ket{{\uparrow}}$ tunnelling to the SET occurs. Here, the electron can tunnel back and forth to the SET; however, if $\ket{{\downarrow}}$ is loaded onto the DD then the tunnel out rate to the SET is much slower (this is schematically shown in Fig.~\ref{fig:intro}c). This results in periods of fast tunnelling (${\ket{{\uparrow}}} {\leftrightarrow} {\ket{0}}$) interspersed with periods of slow tunnelling (${\ket{{\downarrow}}} {\leftrightarrow} {\ket{0}}$). This extra spin state gives rise to the observed super-Poissonian ($\textrm{\textit{FF}}{>}1$) statistics in the counting statistics as it acts as a blocking state~\cite{PhysRevB.91.235413}. We can confirm the presence of a blocking state by looking at the second-order correlation function of the RTS trace, shown in Fig.~\ref{fig:ff_g2_c3}e for three different detuning positions. Near position \raisebox{.5pt}{\textcircled{\raisebox{-.9pt} {2}}} where $\textrm{\textit{FF}}{\approx}0.6$, two-state dynamics and anti-bunching of tunnel events are observed, confirmed by $g^{(2)}(t){<}1$. At the peak of the $\textrm{\textit{FF}}{\approx}1.5$ (position \raisebox{.5pt}{\textcircled{\raisebox{-.9pt} {3}}}), there is clear evidence of the bunching of the electrons for $t{<}0.2$~ms since $g^{(2)}(t){>}1$.

\begin{figure}
\begin{center}
\includegraphics[width=1\columnwidth]{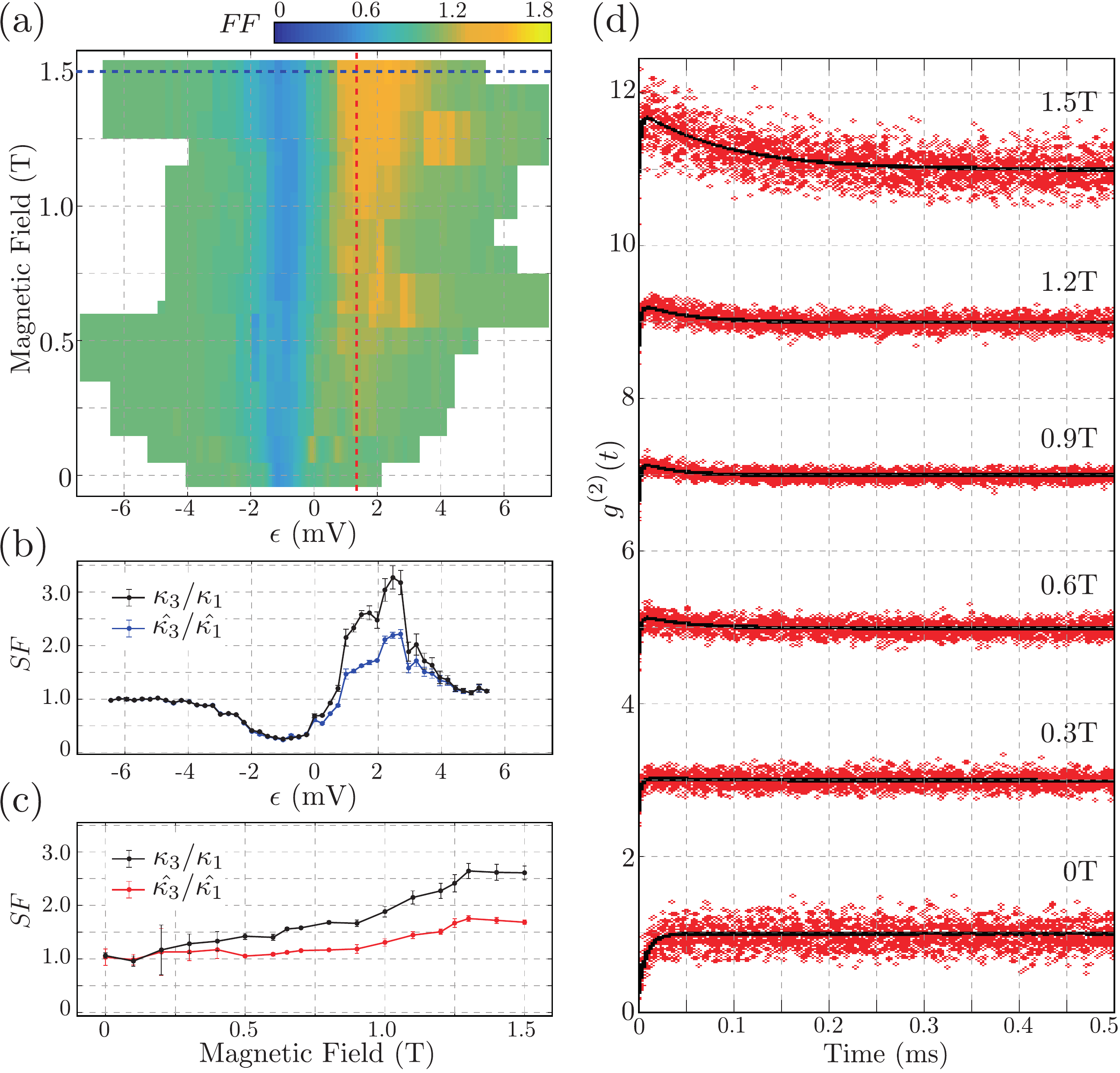}
\end{center}
\caption{{\bf Transition from two- to three-state system.} (a) The magnetic field, $B$ dependence of the measured Fano factor, $FF{=}\hat{\kappa}_2/\hat{\kappa}_1$. The peak can be seen to emerge at low magnetic fields and increase in height as $B$ increases. The data is aligned such that the minima in the Fano factors for different $B$ values are at the same detuning value. (b) The detuning dependence of the calculated and measured normalised skewnes, $SF$; $\kappa_3/\kappa_1$ and  $\hat{\kappa}_3/\hat{\kappa}_1$ at $B{=}1.5$ T (blue dashed line in (a)). The calculated and measured cumulants only differ around the peak in the Fano factor, indicating that the two-state model must be rejected in this detuning regime. (c) $\kappa_3/\kappa_1$ and  $\hat{\kappa}_3/\hat{\kappa}_1$ as a function of $B$ along the detuning position indicated in (a) at $\epsilon{=}1.5$~mV (red dashed line). The cumulants become significantly different above $B{=}0.4$~T, showing that the three state model is required above this magnetic field strength. (d) Selected $g^{(2)}(t)$ traces for different $B$ fields from 0 to 1.5 T at $\epsilon{=}1.5$~mV (offset by 2). As the magnetic field is increased, the bunching of electrons, $g^{(2)}(t){>}1$ becomes more prominent since the ratio of the spin state tunnel rates difference becomes larger. The extent in time of the $g^{(2)}(t){>}1$ region also increases, again, indicative that the ratio between the two spin state tunnel rates becomes larger as the magnetic field is increased. The solid lines are fits to the data using Eq.~\ref{eq:g2calc}.}
\label{fig:bfield}
\end{figure}

We again examine the rf-power dependence on the FCS at high magnetic field in Fig.~\ref{fig:ff_g2_c3}f. The region where $\textrm{\textit{FF}}{>}1$ widens in detuning and decreases in amplitude at larger powers since the effective temperature of the system increases, causing the tunnel rates of the two spin states to become more similar. As before, the dynamics of the system can be described by considering only an effective temperature broadening, indicating that the rf-driving does not give rise to any new dynamics in the system. This is most likely due to the much slower tunnel rates ($\sim 50$~kHz) compared to the rf-driving frequency ($228.6$~MHz). 

\subsection{Intermediate magnetic fields}
\label{sec:Binter}

The condition $\textrm{\textit{FF}}{>}1$ gives an indication of where the two-state (degenerate spin) model cannot be used to describe the system~\cite{bruderer2014}. This is where the effective temperature of the system is too low to resolve the Zeeman split spin states. In Fig.~\ref{fig:bfield}a we plot the \textit{FF} as a function of magnetic field and detuning. The \textit{FF} peak increases in magnitude and width as the magnetic field is increased showing that the magnetic field has a direct effect on the system dynamics. To investigate the transition from a two-state to a three-state system, we measure the normalised skewness, $SF{=}\kappa_3/\kappa_1$ at magnetic field values in the intermediate regime, $g \mu_B B{\sim}k_B T$ or $0{<}B{<}1$~T. 

The two-state system has only two independent cumulants, $\kappa_{1}$ and $\kappa_{2}$, that is, any cumulant $\kappa_{i>2}$ can be written as a function of the preceding cumulants which allows us to determine a statistical test of the system dimension~\cite{bruderer2014}. In this section we test the hypothesis that the DD and rf-SET electron system is classical (in the sense that the Hamiltonian only contains non-zero diagonal elements) and of dimension, $M{=}2$. To this end, we have to measure the first three cumulants, $\{\hat{\kappa}_1,\hat{\kappa}_2,\hat{\kappa}_3\}$. Using the measured first two cumulants, $\hat{\kappa}_1,\hat{\kappa}_2$ we calculate what the third cumulant \textit{would} be for a two-state system,
\begin{equation}
\kappa_3 = \hat{\kappa}_1 + 3 \hat{\kappa}_2 \Big(\frac{\hat{\kappa}_2}{\hat{\kappa}_1} - 1\Big).
\end{equation}
If $\kappa_3{\neq}\hat{\kappa}_3$ then the hypothesis that the system has a dimension $M{=}2$ must be rejected and hence cannot be described by a two dimensional generator, which in our case is $\mathcal{L}_0$.

To investigate the two-state hypothesis we first examine the detuning dependence of $\hat{\kappa}_3$ and $\kappa_3$ at $B{=}1.5$ T, in Fig.~\ref{fig:bfield}b. Importantly, the detuning dependence on $\hat{\kappa}_3$ and $\kappa_3$ shows that the experiment and calculated cumulant only disagree where $\textrm{\textit{FF}}{>}1$ between $\epsilon{\approx}0$~mV and $\epsilon{\approx}4$~mV. This is because the peak in \textit{FF} corresponds to when $\ket{{\uparrow}}$ is above the Fermi level of SET and $\ket{{\downarrow}}$ is below the Fermi level giving different tunnel rates to the SET. Therefore, the three-state model must be used.

\begin{figure}
\begin{center}
\includegraphics[width=1\columnwidth]{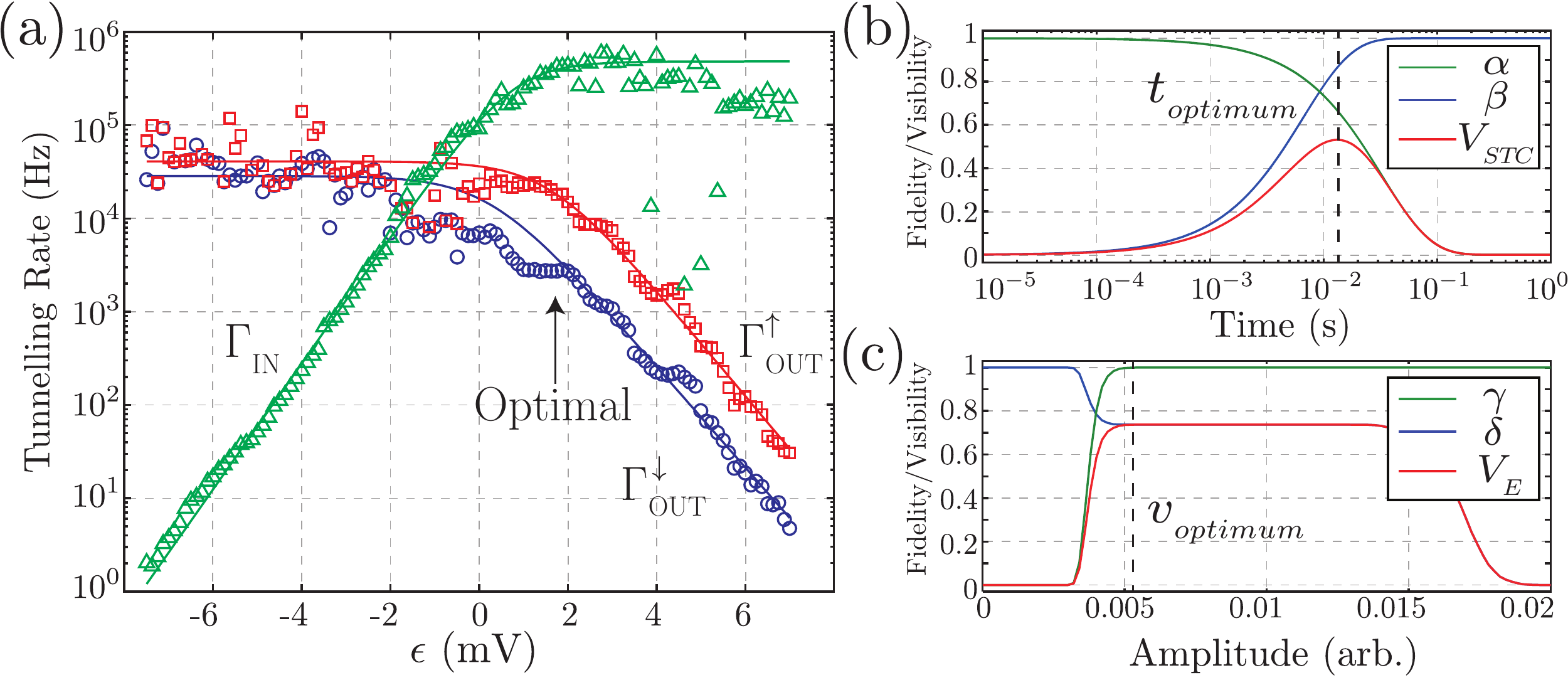}
\end{center}
\caption{{\bf Individual spin tunnel rates for spin-readout.} (a) Individual tunnel rates as a function of detuning, $\epsilon$ at $B{=}2$~T. Two tunnel rates can be observed in the regime where $\textrm{\textit{FF}}{>}1$, which we assign as $\Gamma^{\uparrow}_{\textnormal{OUT}}$ (red squares) and $\Gamma^{\downarrow}_{\textnormal{OUT}}$ (blue circles). A single tunnel in rate is measured since it is the sum of both the individual spin tunnel rates (green triangles). The solid lines are fits to the data using a Fermi-Dirac distribution. Using the data of the tunnel times as well as measuring the signal-to-noise ratio at different powers, we can perform a spin-readout fidelity analysis. (b) The spin-to-charge conversion visibility (red) as well as fidelities, $\alpha$ (blue) and $\beta$ (green) as a function of readout time. (c) The electrical fidelities $\gamma$ (green) and $\delta$ (blue) as well as the electrical visibility (red) as a function of readout threshold. The maximum of these two plots are used to the obtain the optimum readout time and threshold value ($t_{optimum}$ and $v_{optimum}$).}
\label{fig:readout}
\end{figure}

To examine the magnetic field dependence, we take a cut through the \textit{SF} at $\epsilon{=}1.5$~mV shown in Fig.~\ref{fig:bfield}c. The transition from the two-state to three-state system, that is, where $\kappa_3{\neq}\hat{\kappa}_3$ occurs around $B_{\textnormal{tran}}{=}0.4{\pm}0.1$~T. This magnetic field strength, $B_{\textnormal{tran}}$ represents the point where the thermal/power broadening of the rf-SET causes the individual spin state, $\ket{{\uparrow}}$ and $\ket{{\downarrow}}$ tunnel rates to become indistinguishable~\cite{PhysRevLett.111.126803}.

Finally, we plot $g^{(2)}(t)$ in Fig.~\ref{fig:bfield}d for various magnetic field strengths along the same detuning as in Fig.~\ref{fig:bfield}c. As the magnetic field is increased the bunching of tunnelling out events ($g^{(2)}(t){>}1$) becomes more prominent as the difference in energy between the spin-up and spin-down states increases. This is because the blocking spin-down state causes larger periods of no tunnelling. Below $B_{\textnormal{tran}}$ the signature of bunching, that is $g^{(2)}(t){>}1$ disappears and only anti-bunching of electrons can be observed, $g^{(2)}(t){<}1$.

\section{Determining optimal single spin-readout}
\label{sec:readout}

For optimal spin-readout, the electron temperature should be as low as possible to maximise spin-to-charge conversion, which relies on sufficiently different $\Gamma^i_{\textnormal{OUT}}$ for spin-up and -down. The next step is to examine what effect the rf-driving field has on the fidelity of single shot electron spin-readout.

Electron spin-readout fidelities are separated into two processes: electrical visibility and spin-to-charge conversion visibility. Electrical visibility represents the probability of registering a tunnel event (a blip in the detector response) and is governed predominately by the SNR, readout time, and measurement bandwidth of the detector. The spin-to-charge conversion visibility indicates how well the detector is able to distinguish between a tunnel event that is $\ket{{\uparrow}}$ or $\ket{{\downarrow}}$ and depends on the relative tunnel out times of the individual spin states~\cite{buch2013}. We want the tunnel rate of $\ket{{\uparrow}}{\rightarrow}\ket{0}$ to be much greater than $\ket{{\downarrow}}{\rightarrow}\ket{0}$. Therefore, the spin states are positioned such that the $\ket{{\uparrow}}$ chemical potential is above the Fermi level of the SET and that $\ket{{\downarrow}}$ is below the Fermi level. Since we measure the tunnel rates as a function of detuning, we can optimise the readout fidelity over the detuning range and rf-power for a given magnetic field value. An explanation of the various parameters involved in the spin readout fidelity calculation is given in Appendix~\ref{app:readout}.

\subsection{Optimisation of readout time}
\label{sec:Optimtime}

In Fig.~\ref{fig:readout}a we plot the measured tunnel rates obtained from the waiting time distribution of the RTS trace by fitting a double exponential function which gives distinct tunnel out rates (Appendix~\ref{app:temp}) for $\ket{{\uparrow}}$ (higher tunnel rate, red squares) and $\ket{{\downarrow}}$ (lower tunnel rate, blue circles). The tunnel in time corresponds to $\Gamma_{\textnormal{IN}}{=}\Gamma^{\downarrow}_{\textnormal{IN}}{+}\Gamma^{\uparrow}_{\textnormal{IN}}$ and shows only the sum of the two times and therefore only a single exponential can be fit to the data (green triangles). The solid lines are fits to the data using a Fermi-Dirac distribution~\cite{morello2010,buch2013}. The optimum point for spin-to-charge conversion is where the ratio, $\Gamma_{\textnormal{ratio}}{=}\Gamma^{\uparrow}_{\textnormal{OUT}}/\Gamma^{\downarrow}_{\textnormal{OUT}}$ is maximised. From Fig.~\ref{fig:readout}a we can see that $\Gamma_{\textnormal{ratio}}$ does not vary over the detuning range, $\epsilon{>}2$~mV, implying that any point in this region will give the optimal spin-to-charge conversion fidelity. However, the optimum readout time will be faster as $\Gamma^{\uparrow}_{\textnormal{OUT}}$ becomes faster (moving towards negative detuning), meaning that the readout time can be tuned over many orders of magnitude depending on the position in detuning, while maintaining the same spin-to-charge conversion fidelity. Interestingly, the detuning point that gives the fastest readout whilst maintaining the highest spin-to-charge conversion occurs at the peak of the \textit{FF}, denoted by the black arrow in Fig.~\ref{fig:ff_g2_c3}d, which at $B{=}2$~T corresponds to ${\sim}2$~mV. What this means is that by measuring the \textit{FF} as a function of detuning the optimal readout position can be easily found from $\max[FF]$.

\subsection{Optimisation of rf-power}
\label{sec:Optimpower}

Using the data from Fig.~\ref{fig:readout}a we calculate the spin-to-charge conversion visibility ($V_{STC}{=}\alpha{+}\beta{-}1$) and the electrical visibility ($V_{E}{=}\gamma{+}\delta{-}1$) in Fig.~\ref{fig:readout}b and c which are used to obtain the electron spin-readout fidelity. This type of analysis has been reported before~\cite{buch2013} and can used to directly obtain the optimum readout time (Fig.~\ref{fig:readout}b) and the optimal threshold for the tunnel event (Fig.~\ref{fig:readout}c). We now use the same methods to find the optimum spin-readout fidelity for different rf-powers.

\begin{figure}
\begin{center}
\includegraphics[width=1\columnwidth]{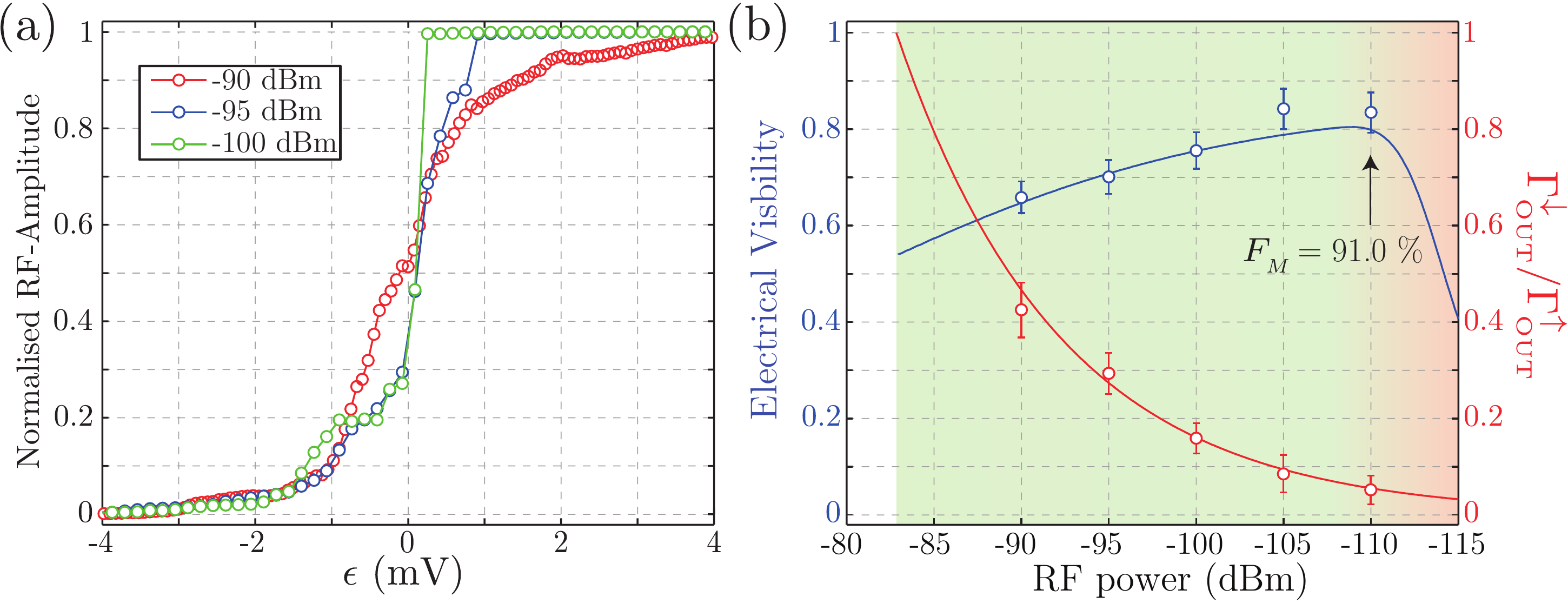}
\end{center}
\caption{{\bf The effect of rf-power on spin-readout fidelity.} (a) The edge of the SET rf-amplitude response across the DD anti-crossing line. As the rf-power is increased the density of states in the SET broadens in detuning. (b) The calculated electrical visibility (blue) and tunnel rate ratio, $\Gamma^{\downarrow}_{\textnormal{OUT}}/\Gamma^{\uparrow}_{\textnormal{OUT}}$ (red) as a function of rf-power to the SET. The data points show the measured values in the experiment and the solid lines are theoretical calculations using parameters obtained from the experiment. There is an optimum rf-power for electron spin-readout, which we calculate here to be $V_E{=}84.2$\%, which, when combined with the spin-to-charge conversion analysis, gives a predicted measurement fidelity, $F_M{=}91.0$\% at -110~dBm.} 
\label{fig:fidelity}
\end{figure}

At higher rf-powers the effective temperature of the system increases. This is confirmed in Fig.~\ref{fig:fidelity}a, where we show the structure of the rf-SET response across the charge transition with the DD. The higher effective tempearture reduces the spin-to-charge conversion fidelity since the $\ket{{\uparrow}}$ and $\ket{{\downarrow}}$ tunnel events become less distinguishable. In Fig.~\ref{fig:fidelity}b we show that as the applied rf-power increases the SNR also increases which gives better electrical fidelity. However, there is a trade off since the tunnel out times of spin-up and -down become more similar as the effective temperature increases. The electrical visibility has three distinct regimes. For small rf-power the SNR becomes too small to accurately register any tunnel event (red region). In the intermediate regime (green), the visibility reaches a maximum and slowly decreases as more rf-power is applied to the rf-SET. The decrease is due to the tunnel out rates becoming too similar. This means the optimum readout time (calculated from spin-to-charge conversion) becomes much shorter and a large number of $\ket{{\uparrow}}$ tunnel events are missed~\cite{buch2013}. When the ratio of the tunnel rates become 1 (large rf-power) then spin-readout becomes impossible since the tunnel events are indistinguishable between $\ket{{\uparrow}}$ and $\ket{{\downarrow}}$ (white region). Therefore, there is an optimum power for spin-readout, which for the device measured here is -110~dBm (effective temperature of $\sim 0.8$~K), which gives a predicted measurement fidelity of $F_M{=}(\alpha\gamma + \beta\delta)/2{=}91.0\%$~\cite{buch2013}.

The electrical fidelity in this device is limited by the fast $\Gamma^{\downarrow}_{\textnormal{IN}}{\approx}250$~kHz which approaches the measurement bandwidth of our data acquisition device. This means some current blips go undetected by the charge sensor. In this work the SNR is large enough to clearly distinguish between the two states (${\sim}40$ at -90~dBm) and in the future the tunnel rates can be easily decreased by having the DD slightly further away from the rf-SET.

\section{Discussion}
\label{sec:disc}

We have investigated the full counting statistics of a single DD coupled to a rf-SET for single shot electron spin-readout. FCS can be used as a tool for probing the system dynamics and elucidating the optimal conditions to maximise electron spin-readout fidelities. We have shown by studying the tunnelling statistics of electrons that the rf-SET can be used to perform single shot spin readout of electrons. 

We examined the spin-readout fidelities by varying the rf-power of the SET and show that there is a clear optimal power that is a compromise between power broadening and SNR. We show that by simply measuring the \textit{FF} as a function of the detuning between the DD and SET the optimal readout position can be easily found from its maximum value. For this device, we calculate a readout fidelity of $91.0$\% and predict that the rf-SET can be used as a charge sensor with fault-tolerant single shot spin-readout fidelities if the tunnel times of the DD to the rf-SET are increased. In summary, we have shown that by directly coupling a DD to a rf-SET and measuring the tunnelling statistics we can optimise the readout fidelities to allow for fault-tolerant single shot spin-readout.

\section{acknowledgements}
We thank M. Bruderer for enlightening discussions. This research was conducted by the Australian Research Council Centre of Excellence for Quantum Computation and Communication Technology (project no. CE110001027) and the US National Security Agency and US Army Research Office (contract no. W911NF-08-1-0527). M.Y.S. acknowledges an ARC Laureate Fellowship.

\appendix
\section{Calculation of cumulants from RTS traces}
\label{app:calc}

In this section, we describe how the RTS traces can be analysed to obtain the cumulants of the distribution, $p(n)$ using FCS. We position the voltage levels on the gates such that the electron can tunnel between the rf-SET and DD. We then wait at this position for $\tau_M{=}100$~s while monitoring the reflected amplitude of the SET. The RTS traces are then sectioned into consecutive windows of a length, $\tau{=}10$~ms for a total of $10,000$ windows. The number of tunnel outs, that is the number of times the RTS traces goes from a low value to a high value (see Fig.~\ref{fig:intro}(c)) per window is then binned into a histogram over the whole RTS trace. An example of the resulting histogram of $p(n)$ is shown in Fig.~\ref{fig:intro}d.

The cumulants of the $p(n)$ can be calculated by first calculating the moments of the distribution. The moments, $\mu_i$ of $p(n)$ are found using,
\begin{equation}
\mu_i = E[p(n)^i],
\end{equation}
where $E[\cdot]$ represents the expectation value (mean) of the distribution. The cumulants can then be found from the recursion formula,
\begin{equation}
\kappa_i = \mu_i - \sum_{j=1}^{i-1} \frac{(i - 1)!}{(j - 1)!(i - j)!} \kappa_j \mu_{i-j}.
\end{equation}
The next RTS trace was then taken by shifting the voltages on the gates along the detuning line, shown by the white line in Fig.~\ref{fig:intro}b and performing the same analysis as above. The distribution of waiting times was calculated from each measured $t_{\textnormal{IN}}$ and $t_{\textnormal{OUT}}$ in Fig.~\ref{fig:intro}c and then binned into a histogram.

\section{Temperature estimations}
\label{app:temp}

The temperature estimations in the main text were found by fitting both the individual cumulants from the FCS analysis and the relative magnitude of the spin tunnel out rates as a function of detuning. The temperature can be found by using the relative magnitude of the tunnel rates if the Zeeman energy, $E_z{=}g \mu_B B$ of the electron spin states are known. The tunnel out rates, $\Gamma^{\downarrow}_{\textnormal{OUT}}$ and $\Gamma^{\uparrow}_{\textnormal{OUT}}$ follow Fermi distributions about the Fermi level of the reservoir,
\begin{equation}
\Gamma^{\downarrow}_{\textnormal{OUT}} = [1 - f(\epsilon - E_z/2)] \Gamma_{\textnormal{OUT}},
\end{equation}
\begin{equation}
\Gamma^{\uparrow}_{\textnormal{OUT}} = [1 - f(\epsilon + E_z/2)] \Gamma_{\textnormal{OUT}},
\end{equation}
where $\Gamma_{\textnormal{OUT}}$ is the maximum tunnel rate. Therefore, the the relative magnitude between the two tunnel out rates is,
\begin{equation}
\frac{\Gamma^{\uparrow}_{\textnormal{OUT}}}{\Gamma^{\downarrow}_{\textnormal{OUT}}} = \frac{1 - f(\epsilon + E_z/2)}{1 - f(\epsilon - E_z/2)}.
\label{eq:tun_rate}
\end{equation}
At far positive detuning when $f(\epsilon + E_z/2){<}1$, Eq.~\ref{eq:tun_rate} is approximately independent of detuning and is given by the ratio $E_z$ to $k_B T$. That is,
\begin{equation}
\lim_{\epsilon\to\infty} \frac{\Gamma^{\uparrow}_{\textnormal{OUT}}}{\Gamma^{\downarrow}_{\textnormal{OUT}}} = \exp{\Bigg(\frac{E_z}{k_B T}\Bigg)},
\end{equation}
such that, after inverting,
\begin{equation}
T = \frac{E_z}{k_B (\ln{\Gamma^{\uparrow}_{\textnormal{OUT}}} - \ln{\Gamma^{\downarrow}_{\textnormal{OUT}}})}.
\end{equation}
The temperature obtained using this method showed good agreement to the cumulants obtained using FCS and as such was used to estimate the temperature of the system. At far positive detuning, $\Gamma^{\uparrow}_{\textnormal{OUT}}/\Gamma^{\downarrow}_{\textnormal{OUT}}{\approx}7$, which gives $T{=}1.4{\pm}0.2$~K for -100~dBm (-95~dBm is $T{=}2.3{\pm}0.3$~K and -90~dBm is $T{=}3.1{\pm}0.4$~K).

\section{Single spin readout parameters}
\label{app:readout}

The electrical visibility is how well the blip in the detector response can be resolved. It is parameterised by two fidelities, $\gamma$ and $\delta$ that correspond to the distributions of the spin-down, $N_{\downarrow}$ and spin-up state, $N_{\uparrow}$,
\begin{align}
\gamma &= 1 - \int_{-\infty}^{v} N_{\downarrow} dV,\\
\delta &= 1 - \int_{v}^{\infty} N_{\uparrow} dV.
\end{align}
Here, $N_{\downarrow}$ is the distribution of the readout trace when there was no blip and $N_{\uparrow}$ when there is a blip present and $V$ is the reflected rf-amplitude. The optimal threshold voltage, $v_{optimum}$ is the value that maximises the separation between the two distributions~\cite{buch2013}. This can be conveniently calculated by maximising the electrical visibility,
\begin{equation}
V_{E} = \gamma + \delta - 1.
\end{equation}

The state-to-charge conversion visibility is calculated by considering a rate equation model of the single electron tunneling to the SET. The two parameters that are used to maximise the probability that the electron tunneling to the SET is a spin-up are $\alpha$ and $\beta$,
\begin{equation}
\alpha = e^{-\frac{t}{t^0_{\textnormal{OUT}}}},
\end{equation}
\begin{multline}
\beta = \frac{1}{T_{\textnormal{OUT}}}\Big[(1 - e^{-\frac{t}{t^0_{\textnormal{OUT}}}}) t^0_{\textnormal{OUT}} t^1_{\textnormal{OUT}}\\ + (e^{-\frac{T_1 + t^1_{\textnormal{OUT}}}{t^1_{\textnormal{OUT}} T_1} t} - 1) T_1 (t^1_{\textnormal{OUT}} - t^0_{\textnormal{OUT}})\Big],
\end{multline}
where $T_{\textnormal{OUT}}{=}T_1 (t^0_{\textnormal{OUT}}{-}t^1_{\textnormal{OUT}}){+}t^0_{\textnormal{OUT}} t^1_{\textnormal{OUT}}$. The fidelity $\alpha$ is the probability that the spin-down electron has not tunneled to the SET and $\beta$ is the probability that the spin-up electron has tunneled to the SET~\cite{buch2013}. The optimal readout time, $t_{optimum}$ is that the time that maximises these two fidelities. This can be found by maximising the state-to-charge conversion visibility,
\begin{equation}
V_{STC} = \alpha + \beta - 1.
\end{equation}

Finally, we define the measurement fidelity as the average probability of correctly identifying the spin-down and spin-up states. This is given by,
\begin{equation}
F_{M} = \frac{F_{\downarrow} + F_{\uparrow}}{2} = \frac{\alpha \gamma + \beta \delta}{2},
\end{equation}
to take into account the effect of the state-to-charge conversion and electrical visibility.

\bibliographystyle{aip}

\end{document}